\documentclass[12pt,a4paper]{article}
\usepackage{amsmath}

\begin{document}

\title{\textbf{\large{Quantum dense coding using three qubits}}}
\vspace{1cm}
\author{Jos\'{e} L. Cereceda\thanks{Electronic mail: jl.cereceda@teleline.es} \\
\textit{C/Alto del Le\'{o}n 8, 4A, 28038 Madrid, Spain}}

\date{May 21, 2001}

\maketitle

\begin{abstract}
We consider a situation in which two parties, Alice and Bob, share a 3-qubit system coupled in an initial maximally entangled, GHZ state. By manipulating locally two of the qubits, Alice can prepare any one of the eight 3-qubit GHZ states. Thus the sending of Alice's two qubits to Bob, entails 3 bits of classical information which can be recovered by Bob by means of a measurement distinguishing the eight (orthogonal) GHZ states. This contrasts with the 2-qubit case, in which Alice can prepare any of the four Bell states by acting locally only on one of the qubits.

\vspace{.5cm}
\noindent
PACS number(s): 03.67.Hk, 03.65.Ud
\end{abstract}

\newpage

In 1992, Bennett and Wiesner [1] characterized the set of 2-qubit states preparable by 1-qubit unitary operators acting on one member of a pair of qubits in an initial Einstein-Podolsky-Rosen (EPR) [2] state, such as the singlet state of two spin-$\frac{1}{2}$ particles. They showed that such a set of states is a proper subset, but not a subspace, of all 2-qubit pure states. It includes, in particular, the four mutually orthonormal Bell states
\begin{align}
|\Phi^{+}\rangle &= (|00\rangle + |11\rangle)/\sqrt{2},  \nonumber  \\
|\Phi^{-}\rangle &= (|00\rangle - |11\rangle)/\sqrt{2},  \nonumber  \\[-.35cm]
&  \\[-.35cm]
|\Psi^{+}\rangle &= (|01\rangle + |10\rangle)/\sqrt{2},  \nonumber  \\
|\Psi^{-}\rangle &= (|01\rangle - |10\rangle)/\sqrt{2},  \nonumber  
\end{align}
where $\{|0\rangle,|1\rangle\}$ is a chosen computational basis of a qubit. The inclusion of the four Bell states (1) in the set of states accessible from an initial EPR state by 1-qubit operations can be exploited to reliably encode a 2-bit message by acting on a single qubit [1]: Suppose that two parties, Alice and Bob, are distantly separated in space but they initially share one of the Bell states, say $|\Phi^{+}\rangle$, where the first qubit is with Alice and the second with Bob. Then, interacting with the first qubit, Alice can transform $|\Phi^{+}\rangle$ into each of the states $|\Phi^{-}\rangle$, $|\Psi^{+}\rangle$, and $|\Psi^{-}\rangle$. So, for example, the unitary operation on the first qubit defined by $|0\rangle \to -|1\rangle$ and $|1\rangle \to |0\rangle$ (that is, state exchange and relative phase shift of $\pi$ between $|0\rangle$ and $|1\rangle$) allows Alice to change the 2-qubit state $|\Phi^{+}\rangle$ to $|\Psi^{-}\rangle$. Application of the appropriate unitary operation on Alice's qubit results in one of the four Bell states (1). Thus Alice can communicate a 2-bit message to Bob by applying one out of four unitary transformations to her qubit, and then sending this single qubit to Bob. Upon reception, Bob performs a joint measurement on the 2-qubit system in the Bell basis, thereby learning which operation Alice applied. This enhances the amount of information transmitted by Alice's single qubit to two bits, doubling the classical maximum capacity of one bit. Partial experimental verification of this ``superdense'' coding protocol has been reported in [3].

In this paper we show how this procedure can be extended to the case in which Alice and Bob have three qubits at their disposal. It is assumed that they initially share a maximally entangled 3-qubit state of the Greenberger-Horne-Zeilinger (GHZ) kind [4]. This state, together with the remaining seven GHZ states, form an orthonormal basis for the Hilbert state space of three qubits (see Eq.\ (2) below). The chief difference with respect to the 2-qubit case is that now Alice cannot transform the initial GHZ state (whichever this might be) into each of the remaining GHZ states by acting only on one qubit. Therefore, in order for Alice to encode a 3-bit message it is necessary that she has access to \textit{two\/} of the qubits. The eight GHZ basis states can be taken to be
\begin{align}
|\psi_1\rangle &= (|000\rangle + |111\rangle)/\sqrt{2},  \nonumber  \\
|\psi_2\rangle &= (|000\rangle - |111\rangle)/\sqrt{2},  \nonumber  \\
|\psi_3\rangle &= (|011\rangle + |100\rangle)/\sqrt{2},  \nonumber  \\
|\psi_4\rangle &= (|011\rangle - |100\rangle)/\sqrt{2},  \nonumber  \\[-.35cm]
&  \\[-.35cm]
|\psi_5\rangle &= (|010\rangle + |101\rangle)/\sqrt{2},  \nonumber  \\
|\psi_6\rangle &= (|010\rangle - |101\rangle)/\sqrt{2},  \nonumber  \\
|\psi_7\rangle &= (|001\rangle + |110\rangle)/\sqrt{2},  \nonumber  \\
|\psi_8\rangle &= (|001\rangle - |110\rangle)/\sqrt{2}.  \nonumber  
\end{align}
By examining (2), it is easily verified that the states $|\psi_2\rangle$, $|\psi_3\rangle$, and $|\psi_4\rangle$ are accessible from $|\psi_1\rangle$ by unitary transformations on the first qubit. However, starting from $|\psi_1\rangle$, one cannot make the states $|\psi_5\rangle$, $|\psi_6\rangle$, $|\psi_7\rangle$, or $|\psi_8\rangle$ by unitarily operating only on the first qubit. Similarly, manipulating the first qubit in the state $|\psi_8\rangle$, one can obtain the states $|\psi_5\rangle$, $|\psi_6\rangle$, and $|\psi_7\rangle$, but not $|\psi_1\rangle$, $|\psi_2\rangle$, $|\psi_3\rangle$, or $|\psi_4\rangle$.

The impossibility to generate all the basis states by operating locally on one of the three qubits is not a specific feature of the GHZ basis (2) but, rather, it is a property common to all the orthonormal bases spanning the state space of the 3-qubit system. A more general example illustrating this fact is provided by the following orthonormal basis
\begin{align}
|\phi_1\rangle &= (|000\rangle + |001\rangle + |010\rangle + |011\rangle + |100\rangle + |101\rangle + |110\rangle + |111\rangle)/2\sqrt{2},  \nonumber  \\
|\phi_2\rangle &= (|000\rangle + |001\rangle + |010\rangle + |011\rangle - |100\rangle - |101\rangle - |110\rangle - |111\rangle)/2\sqrt{2},  \nonumber  \\
|\phi_3\rangle &= (|000\rangle + |001\rangle - |010\rangle - |011\rangle - |100\rangle - |101\rangle + |110\rangle + |111\rangle)/2\sqrt{2},  \nonumber  \\
|\phi_4\rangle &= (|000\rangle + |001\rangle - |010\rangle - |011\rangle + |100\rangle + |101\rangle - |110\rangle - |111\rangle)/2\sqrt{2},  \nonumber  \\[-.35cm]
&   \\[-.35cm]
|\phi_5\rangle &= (|000\rangle - |001\rangle + |010\rangle - |011\rangle - |100\rangle + |101\rangle + |110\rangle - |111\rangle)/2\sqrt{2},  \nonumber  \\
|\phi_6\rangle &= (|000\rangle - |001\rangle + |010\rangle - |011\rangle + |100\rangle - |101\rangle - |110\rangle + |111\rangle)/2\sqrt{2},  \nonumber  \\
|\phi_7\rangle &= (|000\rangle - |001\rangle - |010\rangle + |011\rangle - |100\rangle + |101\rangle - |110\rangle + |111\rangle)/2\sqrt{2},  \nonumber  \\
|\phi_8\rangle &= (|000\rangle - |001\rangle - |010\rangle + |011\rangle + |100\rangle - |101\rangle + |110\rangle - |111\rangle)/2\sqrt{2}.  \nonumber  
\end{align}
From Eq.\ (3), it can be seen that the states obtainable from, say $|\phi_5\rangle$ by unitary transformations acting on the first qubit are (apart from $|\phi_5\rangle$ itself) $|\phi_6\rangle$, $|\phi_7\rangle$, and $|\phi_8\rangle$, whereas the states $|\phi_1\rangle$, $|\phi_2\rangle$, $|\phi_3\rangle$, and $|\phi_4\rangle$ are not. On the other hand, it is the case that only the state $|\phi_2\rangle$ ($|\phi_4\rangle$) is accessible from $|\phi_1\rangle$ ($|\phi_3\rangle$) by acting on the first qubit. For a general orthonormal basis $\{|\Phi_i\rangle\}$, with $\langle\Phi_i|\Phi_j\rangle = \delta_{ij}$, $i,j=1,2,\ldots,8$, it will be definitely impossible to change an initial basis state $|\Phi_i\rangle$ to any of the states $|\Phi_j\rangle$, $i\neq j$, by unitarily operating only on one qubit.

Let us return to the GHZ basis (2) and suppose that Alice and Bob initially share one of the GHZ states, say $|\psi_1\rangle$. It will also be assumed that Alice is initially in possession of two of the qubits, say qubits 1 and 2, whereas qubit 3 is with Bob.  Now Alice can effectively change the state $|\psi_1\rangle$ to any of the GHZ sates (2) by applying an appropriate unitary operation on her two qubits. In matrix form, the eight pertinent 2-qubit unitary operations are (written in the $\{|00\rangle,|01\rangle,|10\rangle,|11\rangle\}$ basis)
\begin{align}
U_{1\to 1} &= \begin{pmatrix}
1 &  &  &  \\  & 1 &  &  \\  &  & 1 &  \\  &  &  & 1
\end{pmatrix}, \quad
U_{1\to 2} = \begin{pmatrix}
1 &  &  &  \\  & 1 &  &  \\  &  & -1 &  \\  &  &  & -1
\end{pmatrix}, \quad
U_{1\to 3} = \begin{pmatrix}
 &  & 1 & 0 \\  &  & 0 & 1 \\ 1 & 0 &  &  \\ 0 & 1 &  & 
\end{pmatrix},  \nonumber   \\[.2cm]
U_{1\to 4} &= \begin{pmatrix}
 &  & 1 & 0 \\  &  & 0 & 1 \\ -1 & 0 &  &  \\ 0 & -1 &  & 
\end{pmatrix}, \quad
U_{1\to 5} = \begin{pmatrix}
0 & 1 &  &  \\ 1 & 0 &  &  \\  &  & 0 & 1 \\  &  & 1 & 0
\end{pmatrix}, \quad
U_{1\to 6} = \begin{pmatrix}
0 & -1 &  &  \\ 1 & 0 &  &  \\  &  & 0 & -1 \\  &  & 1 & 0
\end{pmatrix},   \nonumber     \\[-.25cm]
&   \\[-.25cm]
U_{1\to 7} &= \begin{pmatrix}
 &  &  & 1 \\  &  & 1 &  \\  & 1 &  &  \\ 1 &  &  & 
\end{pmatrix}, \quad
U_{1\to 8} = \begin{pmatrix}
 &  &  & 1 \\  &  & 1 &  \\  & -1 &  &  \\ -1 &  &  & 
\end{pmatrix},  \nonumber
\end{align}
where $U_{1\to j}$ is the transformation which changes the initial state $|\psi_1\rangle$ to $|\psi_j\rangle$. Thus Alice can transmit a 3-bit message by applying one out of the eight transformations (4) to her two qubits, and then sending this pair of qubits to Bob. Once Bob is in possession of the three qubits, he performs a joint measurement on them in the GHZ basis to determine which operation Alice applied.

It is to be noted that the coding scheme just described for three entangled qubits is not as ``dense'' as in the original 2-qubit protocol [1], since in the 3-qubit case Alice can encode, on average, 1.5 bits of information per transmitted qubit, which is less than 2 bits per transmitted qubit achieved in the 2-qubit case (although, of course, the total amount of information transmitted by Alice is greater in the 3-qubit case (3 bits) than in the 2-qubit case (2 bits) \footnote{
It must be noticed that, in any case, the capacity of dense coding is optimized when the sender prepares the basis signal states with equal \textit{a priori\/} probabilities.}). As we have seen, this is a direct consequence of the fact that, starting from an initial 3-qubit basis state, it is not possible to prepare all eight orthonormal states in the basis by unitarily interacting only with one of the qubits.

To conclude, it will be noted that, at least theoretically, it is possible to unambiguously identify any of the GHZ states (2) by means of a quantum network involving only controlled-{\small NOT} gates and single-qubit operations [5]. As shown diagrammatically in Fig.\ 1, the three input qubits in the GHZ superposition become \textit{disentangled\/} at the output, ending up in the product state $|ijk\rangle$, so that the GHZ measurement reduces to three single-qubit measurements. As can easily be checked, the overall unitary evolution induced by the network on the GHZ states (2) is given by [5]
\begin{equation}
(1/\sqrt{2})(|0jk\rangle + |1\bar{j}\bar{k}\rangle) \to |0jk\rangle,  \tag{5a}
\end{equation}
and
\begin{equation}
(1/\sqrt{2})(|0jk\rangle - |1\bar{j}\bar{k}\rangle) \to |1jk\rangle,  \tag{5b}
\end{equation}
where $\bar{j}$ and $\bar{k}$ denote the negation of $j$ and $k$, respectively.

A practical experimental scheme for identifying two of the GHZ states using polarizing beam splitters and half-wave plates, was presented in [6]. Moreover, experimental generation and observation of (polarization) GHZ entanglements between three spatially separated photons has recently been reported in [7].

\begin{figure}[ttt]
    
\begin{picture}(250,125)(-55,0)
\thicklines
\put(99,104){1}    
\put(92,100){\line(1,0){43}}
\put(130,100){\circle*{7}}
\put(130,100){\line(1,0){35}}
\put(165,100){\circle*{7}}
\put(165,100){\line(1,0){30}}
\put(195,87.5){\framebox(25,25){\large\textit{\textbf{H}}}}
\put(220,100){\line(1,0){15}}
\put(241,97){$|0\rangle,|1\rangle$}

\put(99,69){2}
\put(95,65){\line(1,0){140}}
\put(165,65){\circle{14}}    
\put(241,62){$|0\rangle,|1\rangle$}
\put(165,100){\line(0,-1){42}}

\put(99,34){3}
\put(92,30){\line(1,0){143}}
\put(130,30){\circle{14}}
\put(241,27){$|0\rangle,|1\rangle$}
\put(130,100){\line(0,-1){77}}

\put(70,65){\oval(50,90)}
\put(54,70){{\large GHZ}}
\put(52,56){{\large states}}

\end{picture}
\vspace{-.7cm}
\caption{\small{Simple quantum network effecting a GHZ measurement.}}
\vspace{.3cm}

\end{figure}
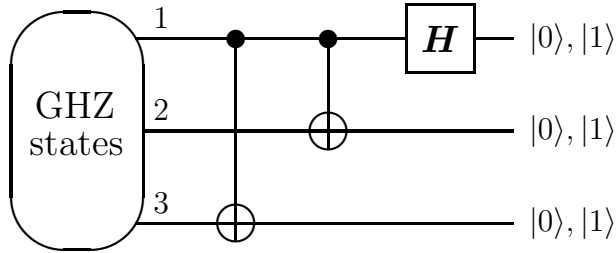

\newpage

\textit{Note added} --- After this work was completed, I learned of a paper by Gorbachev \textit{et al.} [8] which contains similar results (see, specifically, subsection 3.2 of [8]). On the other hand, very recently, Hao \textit{et al.} [9] have proposed an alternative dense coding scheme using the GHZ state in which a \textit{third\/} party can control the quantum channel between Alice and Bob via local measurement and classical communication.

\vspace{.6cm}

\center

\end{document}